\documentclass[amsmath,superscriptaddress,aps,prl,twocolumn]{revtex4}
\usepackage[colorlinks,linkcolor=blue,anchorcolor=blue,citecolor=blue,urlcolor=blue]{hyperref}
\usepackage{amsmath}
\usepackage{amssymb}
\usepackage{graphicx}
\usepackage{color}
\usepackage{bm}
\usepackage{float}
\begin{document}
\title{Thermodynamics and fluctuations in finite-time quantum heat engines under reservoir squeezing}
\author{Yang Xiao}
\affiliation{Department of Physics, Nanchang University,
Nanchang, $330031$, China}

\author{Dehua Liu}
\affiliation{Department of Physics, Nanchang University,
Nanchang, $330031$, China}
\author{Jizhou He}
\affiliation{Department of Physics, Nanchang University,
Nanchang, $330031$, China}
\author{Wu-Ming Liu}
\email{wmliu@iphy.ac.cn}
\affiliation{Beijing National Laboratory for Condensed Matter Physics, Institute of Physics, Chinese Academy of Sciences, Beijing $100190$, China}

\author{L.-L Yan}
\affiliation{School of Physics, Zhengzhou University, Zhengzhou $450001$, China}

\author{Jianhui Wang}
\email{wangjianhui@ncu.edu.cn}
\affiliation{Department of Physics, Nanchang University,
Nanchang, $330031$, China}
\affiliation{State Key Laboratory
of Surface Physics and Department of Physics, Fudan University,
Shanghai $200433$, China}

\begin{abstract} 
We investigate the thermodynamics and fluctuations of a finite-time quantum Otto engine alter-
natively driven by a hot squeezed and a cold thermal reservoir.
We show that   reservoir squeezing   significantly enhances the performance by increasing the thermodynamic efficiency and the power,  and  enables higher stability by   decreasing the relative power fluctuations and speeding up the convergence of quantum efficiency to its most probable value. We also demonstrate the  counterintuitive result that the efficiency can be larger than the Otto limit in the finite-time operation.     Experimental demonstration of this quantum heat engine can be available, based on
 a single-electron spin pertaining to a trapped $^{40}$Ca$^+$
ion \cite{Grb19}.  We provide a general framework for reliably studying  the  finite-time nanoengine in finite time operation which accounts for quantum friction and coherence, deriving important insights into the novel thermodynamic behaviors beyond the classical thermal machines.   
\end{abstract}
\maketitle
\date{\today}
Quantum heat engines have become a laboratory re-
ality, notably the recent experiments realizing quantum
Otto heat engines on nuclear magnetic resonance \cite{RJ19, Ser19}  and  nitrogen-vacancy centers in diamond \cite{JK19}. These thermal machines wherein,  apart from the working substance, the reservoirs may be finite-dimensional and thus non-thermal \cite{Cam16,Zhang14}, have access to nano-scale open systems in which   quantum effects manifest themselves,  such as  coherence \cite{Cam19,Nie18, AW16,KM16, Scu11,Bra17,Rah12,RA15,KE18,Fra20,AG17,Guff19,Scu03},  entanglement \cite{Zhang07,Wangwu19, Huang09,Fun13,Wang08},   correlations \cite{JO02,LJ11,JJ13,Chen11},  quantum measurements \cite{ Hor14,Bra15, Kam16, Cha21, Su21}, and squeezing \cite{Aba14,Kla17,Sin20,You18,Manz18,Assis20,BL22}. The quantum engines in  the presence of these additional freedoms may outperform their classical counterparts \cite{Korzekwa16,Per15,Nie16,RD09,Ber17,And19,LB19,CE17,CE18,MPer15,wang19,xiao18}.
This constitutes one of the central issues in   quantum thermodynamics. 

For microscopic systems, heat and work are no longer deterministic \cite{Sek10,Bou21, Sei12, Smi18,Qian10} as is the case for macroscopic systems. As a result, the efficiency and power for quantum heat engines are stochastic, and both of them are fluctuating.  The power
fluctuations, together with the efficiency fluctuations \cite{Hol121,Lut20}, as a
limiting factor for the practical usefulness in heat engines, measure the machine stability \cite{ Bou21}. Ideally, the quantum heat engine should have  high efficiency (small entropy production), large power, and  small fluctuations for these thermodynamic variables measuring performance. Strong emphasis has been put
on the finite-time thermodynamics of the quantum heat engines, and in particular on fluctuations of power and efficiency \cite{Ver14,Xiao21, Jiang15,Lut20,Lut21, Jiao21,wang18,Jie22}.

Unlike the previous studies considering
nanoengines \cite{RJ19,Ser19,Aba14,Kla17,wang19,xiao18} where quasistatic and local-equilibrium  approximations were required and thus some quantum effects were tied to ignoring, we  develop a formalism for analyzing  the performance and stability for quantum heat engines by overcoming these limitations. We show that both efficiency and power are enhanced by reservoir squeezing with the advantage of decreasing fluctuations of efficiency and power, which is the generic case for finite-time cyclic heat engines driven by non-thermal reservoirs. The result that the efficiency can be enhanced by speeding up the machine even in the absence of squeezing is in stark contrast to previous reports \cite{Guff19,Kos14,Roc20,Sei18}. In particular, we find the counter-intuitive result that the  efficiency can beat the Otto limit when and only when the unitary driving proceeds in finite time.  The result relies only on purely quantum origin and it would not hold when either unitary driven stroke or thermal-contact process satisfies the quasi-static limit.

\begin{figure}
\includegraphics[width=3in]{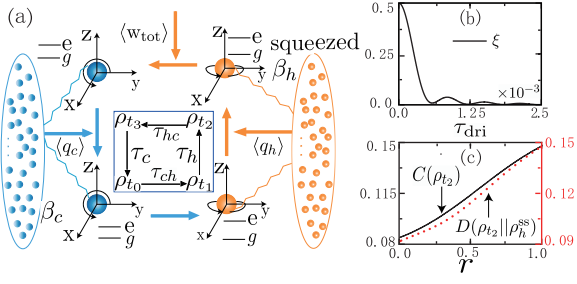}
\caption{ (a) Illustration of  a spin-$1/2$ system  operating
with a quantum Otto cycle alternatively driven by a hot squeezed and
a cold thermal bath.   The  system states   at times $t=t_i$  with $i=0, 1, 2, 3$ are denoted by $\rho_{t_i}$, where $\rho_{t_i}$ are the initial states of the four strokes in a cycle, respectively. 
In each cycle the machine  produces the total work $-\langle\mathrm{w_{tot}}\rangle$
by absorbing average heats from the hot and cold baths, $\langle
q_{h}\rangle$ and $\langle q_{c}\rangle$, where $\langle
q_{h}\rangle=-\langle \mathrm{w_{tot}}\rangle-\langle q_{c}\rangle$
due to the energy conservation. {(b) Transition probability as a function of driving time $\tau_\mathrm{dri}$. (c) Coherence and Kullback-Leibler divergence  at $t=t_2$ with  $\tau_h=0.2$ and  $\tau_{\mathrm{dri}}=5\times10^{-4}$.  The parameters are {$\hbar=1$,} $\omega_c/2\pi= 1000$,
$\omega_h/2\pi=2250$,  $\beta_c=2/(\hbar\omega_c)$,   $\beta_h=1/(\hbar\omega_h)$, $\tau_c=3$, and $\gamma_c=\gamma_h=3$.}} \label{model}
\end{figure}

We consider a quantum Otto engine cycle working between a hot squeezed and a cold thermal bath [see Fig. \ref{model}(a)]. This engine cycle consists of two unitary and two isochoric strokes.
 Firstly, unitary compression from state $\rho_{t_0}$ to $\rho_{t_1}$ with $t_0=0$:  the energy gap is  enlarged by a spin$-1/2$ system with
Hamiltonian
$ H_{{ch}}(t)=\frac{\hbar\omega(t)}{2}\{\cos[\pi t/(2\tau_{{ch}})]\sigma_{x}+\sin[\pi t/(2\tau_{{ch}})]\sigma_{z}\}, $ 
where $\omega(t)=\omega_{{c}}(1-{t}/\tau_{{ch}})+\omega_{h}({t}/{\tau_{{ch}}})$ with $\tau_{{ch}}=t_1$ and $0\le t\le\tau_{ch}$, and  $\sigma_{x,y,z}$ are the Pauli matrices.   The driven Hamiltonian does not commute at different times, generating  quantum coherence  in the energy basis of the system.
Secondly, isochoric heating from state $\rho_{t_1}$ to $\rho_{t_2}$: the system
is weakly coupled to a hot squeezed  reservoir at inverse
temperature $\beta_{h}$ during time
duration $\tau_{h}$ with $\tau_{h}=t_2-t_1$, while its Hamiltonian
keeps a constant as $H_{h}(t)=H_{ch}(t_1)=\hbar
\omega_{h}\sigma_{z}/2$. Thirdly, {unitary expansion} from state
$\rho_{t_2}$ to $\rho_{t_3}$: the driven Hamiltonian $H_{hc}(t_3-t)=H_{ch}(t)$
is realized by reversing the protocol used in the unitary
compression, such that the expansion Hamiltonian takes the
same time as the compression Hamiltonian, namely,
$\tau_{\mathrm{dri}}=\tau_{hc}=\tau_{ch}$. Lastly, {isochoric
cooling} from state $\rho_{t_3}$ to $\rho_{t_3+\tau_{c}}$: the system is weakly
coupled with a cold thermal reservoir at inverse temperature $\beta_{c}$ in
time period $\tau_{c}$, and its Hamiltonian is kept constant at
$H_{c}(t)$=$H_{ch}(0)=\hbar\omega_{c}\sigma_{x}/2$. These times $\tau_h, \tau_c$ and $\tau_{\mathrm{dri}}$ set the total cycle period $\tau_{\mathrm{cyc}}$, namely, $\tau_{\mathrm{cyc}}=\tau_h+\tau_c+2\tau_{\mathrm{dri}}$.  We consider the
machine working in the limit cycle \cite{Kos17} where a periodic
steady state is achieved with all  the periodic variables.

{The dynamics of a quantum system weakly coupled to a heat reservoir of inverse temperature $\beta$ can be described by quantum master equation in Lindblad form \cite{Bre02, Wei12}
\begin{equation}
    \frac{d\rho}{dt}=-\frac{i}{\hbar}[H,\rho_t]+ \mathcal{L}_D(\rho_t), \label{rhot}
\end{equation}
where $\mathcal{L}_D$ is the Lindblad super operator describing heat dissipation  responsible for driving the system  to the steady state where $\rho_t=\rho_t^{ss}=e^{-\beta H}/\mathrm{Tr (e^{-\beta H})}$. The dynamics  of the system during a unitary stroke where no heat is exchanged is given by  $\frac{d\rho}{dt}=-\frac{i}{\hbar}[H,\rho_t]$ \cite{Ser19,Kos17}. The system states   at the respective ends of the two driven  strokes are then  
$
\rho_{t_1}=U_{ch}\rho_{t_0}U_{ch}^{\dag}$and $
\rho_{t_3}=U_{hc}\rho_{t_2}U_{hc}^{\dag},
$
where $U_{ch}=
\mathcal{T}_{>}\exp\{-\frac{i}{\hbar}\int_{t_0}^{t_1}dt
H_{ch}(t)\}$ and $U_{hc}=
\mathcal{T}_{>}\exp\{-\frac{i}{\hbar}\int_{t_2}^{t_3}dt
H_{hc}(t)\}$,  with the time-ordering operator $\mathcal{T}_{>}$. Also, it is noted that in an isochoric where  the system Hamiltonian is fixed,  the dynamics (\ref{rhot})  becomes ${d\rho_t}/{dt}=\mathcal{L}_D(\rho_t)$, where $\rho_t$ is replaced by $\rho_t^{\mathrm{sq}}=\hat{\mathcal{S} }(r)\rho_t\hat{\mathcal{S} }^\dag (r)$ with $\hat{\mathcal{S} }(r)=\mathrm{exp}(r^*\sigma_{-}-r\sigma_{+})$ being dependent on both squeezing parameter $r$ and  $\sigma_\pm=(\sigma_x\pm i\sigma_y)/2$ in the presence of reservoir squeezing.  Hence,  the
quantum Lindblad equation determines  density matrices $\rho_{t_i} (i=0, 1, 2, 3)$  for the spin system [See Supplementary Material (SM), Sec. I\cite{sup}]}.

Based on the aforementioned dynamical description, we 
 derive the  average work $\langle \mathrm{w}_{\mathrm{tot}}\rangle $,  average injection $\langle q_h\rangle$, and   work fluctuations, $\delta \mathrm{w^2_{tot}}=\langle\mathrm{w^2_{tot}}\rangle-\langle \mathrm{w_{tot}}\rangle^{2}$, can be given (see SM   Sec. II \cite{sup}) by  
\begin{eqnarray}
-\langle\mathrm{w}_\mathrm{tot}\rangle&=&\hbar(\omega_{h}-\omega_{c})(\langle n_{t_2}\rangle-\langle n_{t_0}\rangle)\nonumber\\
&+&2\hbar\xi(\omega_{c}\langle n_{t_2}\rangle+\omega_{h}\langle n_{t_0}\rangle)\nonumber\\
&-& 2\hbar \omega_{h}\zeta_{ch}-2\hbar \omega_{c}\zeta_{hc},\label{wtot}\\
\langle q_{h}\rangle&=&\hbar \omega_{h}[\langle n_{t_2}\rangle+\langle
n_{t_0}\rangle(2\xi-1)-2\zeta_{ch}], \label{qh}\\
\delta \mathrm{w^2_{tot}}&=&\hbar^2 \omega_{h}^{2}\{\frac{1}{2}-\langle n_{t_2}\rangle^2-[\langle n_{t_0}\rangle(1-2\xi)+2\zeta_{ch}]^2\}\nonumber\\
&+&\hbar^2 \omega_{c}^{2}\{\frac{1}{2}-\langle n_{t_0}\rangle^2-[\langle n_{t_2}\rangle(1-2\xi)+2\zeta_{hc}]^2\}\nonumber\\
&+&\hbar^2 \omega_{c}\omega_{h}\{2\langle n_{t_0}\rangle[\langle n_{t_0}\rangle (1-2\xi)+2\zeta_{ch}]\nonumber\\
&+&2\langle n_{t_2}\rangle[\langle
n_{t_2}\rangle(1-2\xi)+2\zeta_{hc}]+2\xi-1 \}. \label{wfluc}
\end{eqnarray}
where we have used $\langle n_{t_i}\rangle:=\mathrm{Tr}(\rho_{t_i}H)/(\hbar\omega_i)$ with $\omega_{h}=\omega_{2}$ and $\omega_c=\omega_{0}$ to denote the average populations at times $t=t_i$. Here, 
$\zeta_{ch}:=-\mathrm{Re}[U_{ch}^{{gg}}
\rho_{t_0}^{mn} U_{ch}^{{eg}\dag}]$, 
$\zeta_{hc}:=-\mathrm{Re}[U_{hc}^{{gg}}\rho_{t_2}^{ij}  U_{hc}^{{eg}\dag}]$, and $\xi\equiv |\langle n(t_0)|U_{ch}|m(t_1)\rangle|^{2}=|\langle i(t_2)|U_{hc}|j(t_3)\rangle|^{2} $ with $U_{ch,hc}^{mn}=\langle m(t_{1,3})|U_{ch,hc}|n(t_{0,2})\rangle$ ( $m,n,i,j=e,g$).
{The Hamiltonian-dependent parameter $\xi$ that denotes  the level transition probability during expansion or compression for the spin system  decreases with increasing driving time $\tau_{dri}$, though not monotonically,  and it tends to be zero when the time is long enough to satisfy quantum adiabatic condition, as shown in Fig. \ref{model}(b)}.   

{The parameters $\zeta_{hc}$ and $\zeta_{ch}$ in Eq.
(\ref{wtot}) are associated with the quantum coherence, of which the amount 
 is quantified by the  relative entropy of coherence \cite{Bau14}:
$C(\rho)=S[\mathcal{E}(\rho)]-S(\rho)~(\mathrm{with}~ S(\rho)=-\mathrm{Tr} [\rho \mathrm{ln}  \rho]).
$
Here $\mathcal{E}(\cdot)=\sum_n\Pi_n(\cdot)\Pi_n$ is the dephasing map by removing all coherence in the energy basis, with 
$\Pi_{n}:=|n\rangle\langle n|$ being the propagators of $\rho$. The residual coherence at the end of the hot isohcore, $C(\rho_{t_2})$,   decreases as $r$ increases, as shown in Fig. \ref{model}(c),  where the divergence $D(\rho_{t_2}||\rho_h^\mathrm{ss})=\mathrm{Tr}[\rho_{t_2}(\ln\rho_{t_2}-\ln\rho^\mathrm{ss}_h)]$ with $\rho_h^{\mathrm{ss}}=\rho_{t_2}^{\mathrm{ss}}$, measuring how far the system at the end of the isochore deviates from the steady state, also decreases with increasing $r$ as it should.}  Quantum coherence generated during the two unitary driven strokes  are correlated
for   incomplete thermalization.
 To reveal such  a dynamical interference effect on the thermodynamic
quantities of the machine, our quantum heat engine is compared with
an alternative cycle, where a full dephasing operation \cite{Cam19}
is performed to completely remove all coherence after the hot
isochore with any value of thermal-contact time $\tau_h$. In what
follows we use the superscript ``deph'' to describe the quantities
corresponding to the dephased engine cycle.

\begin{figure} 
\includegraphics[width=3.65cm]{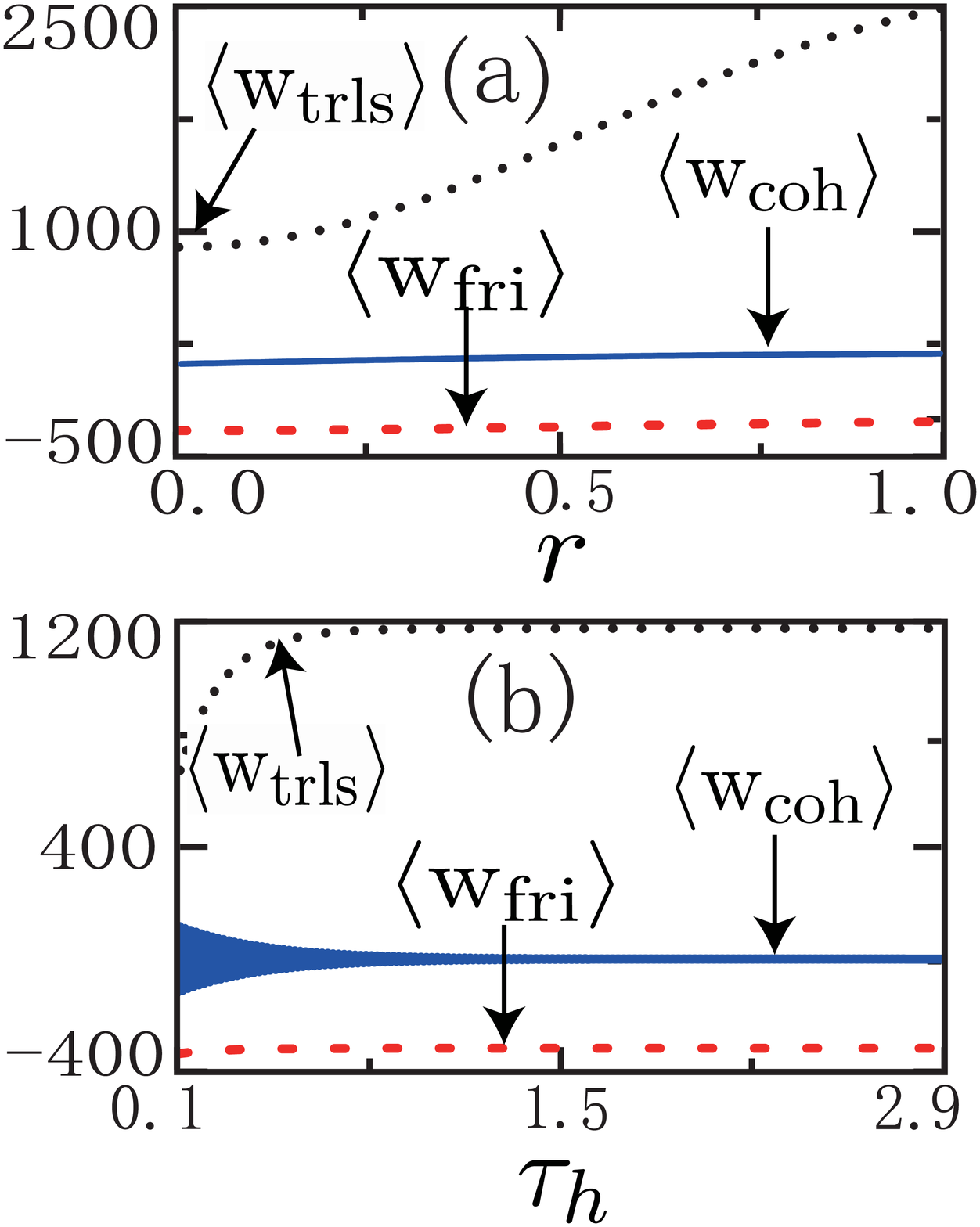}
\includegraphics[width=4.1cm]{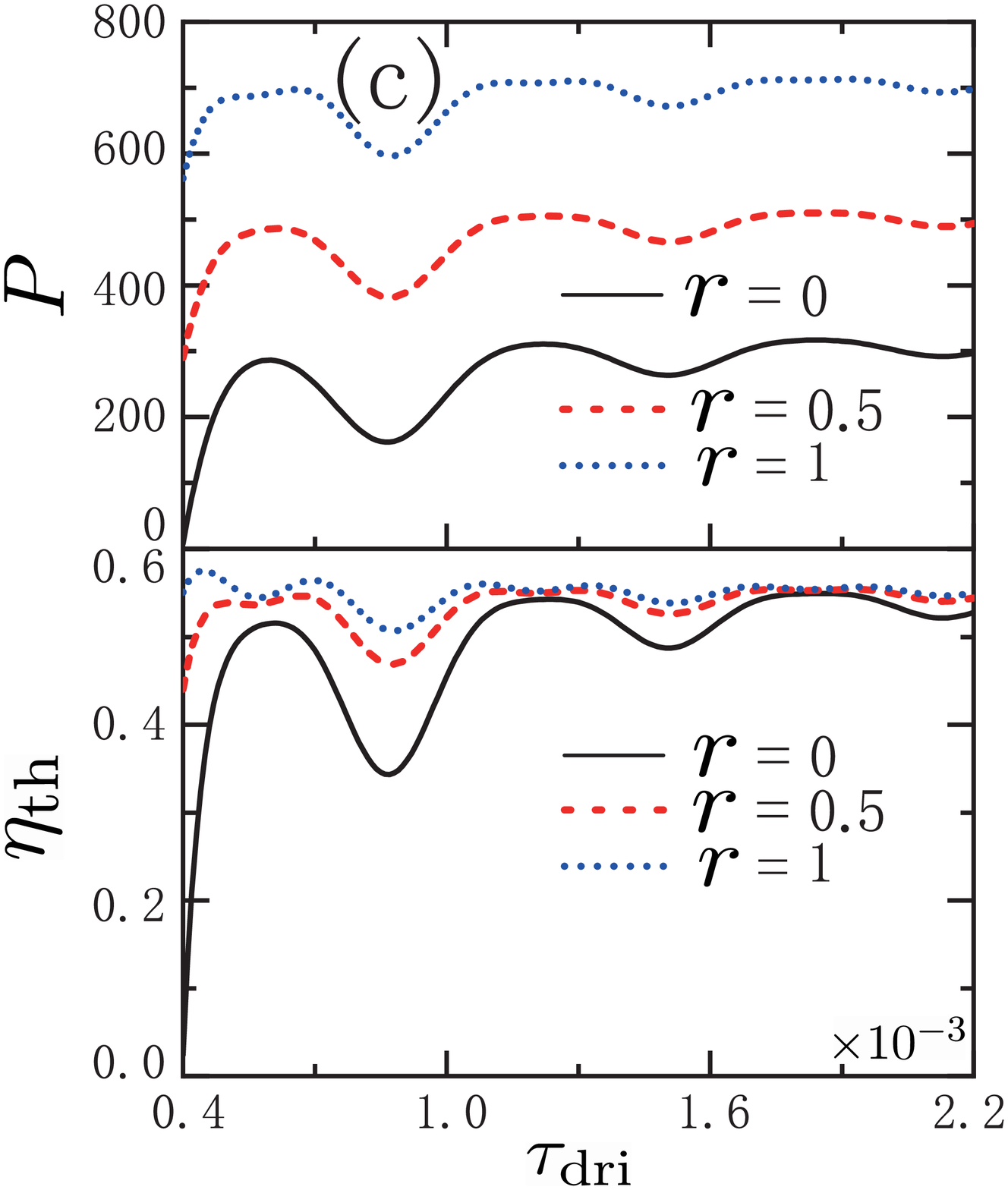}
\includegraphics[width=4.1cm]{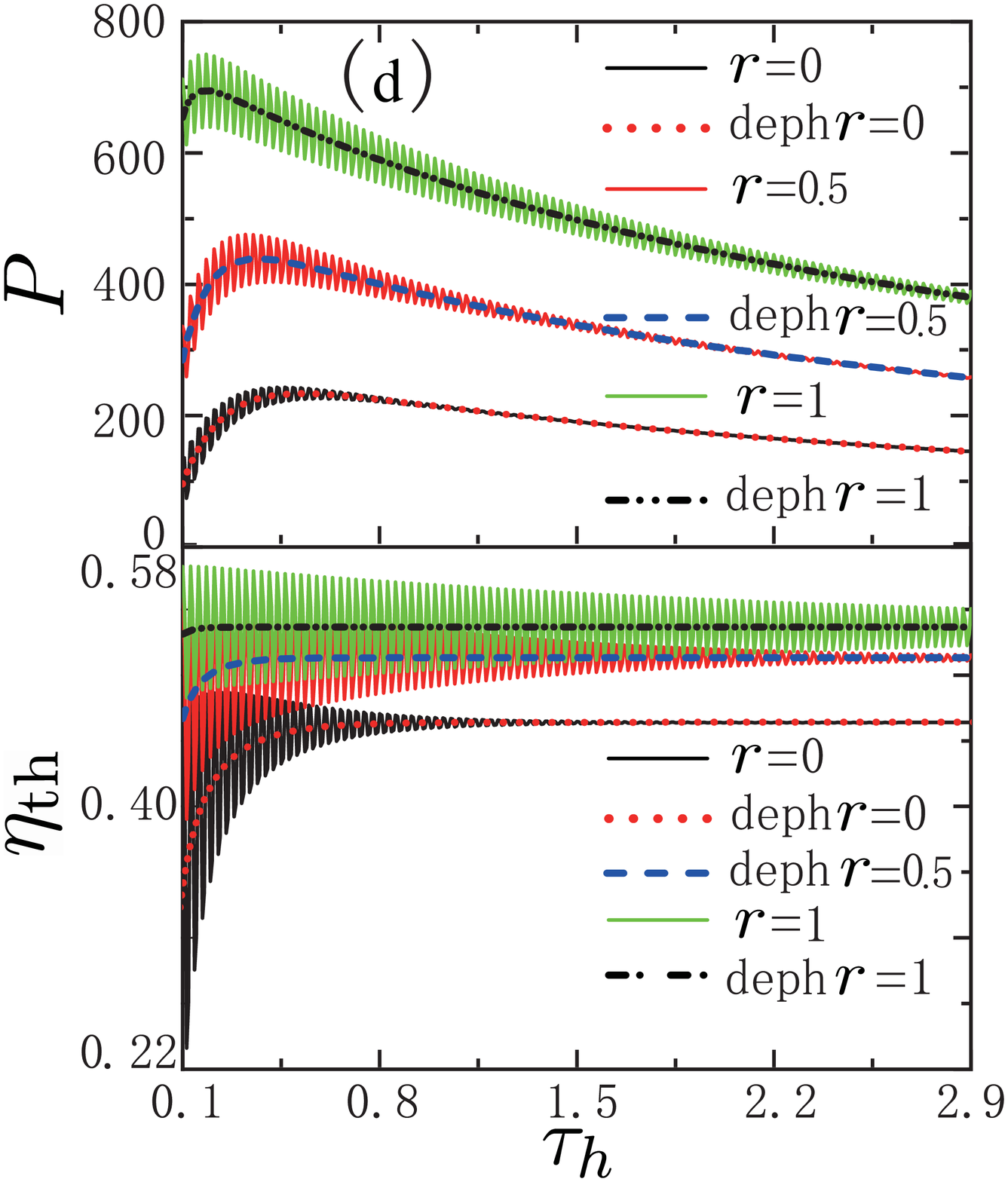}
\includegraphics[width=4.1cm]{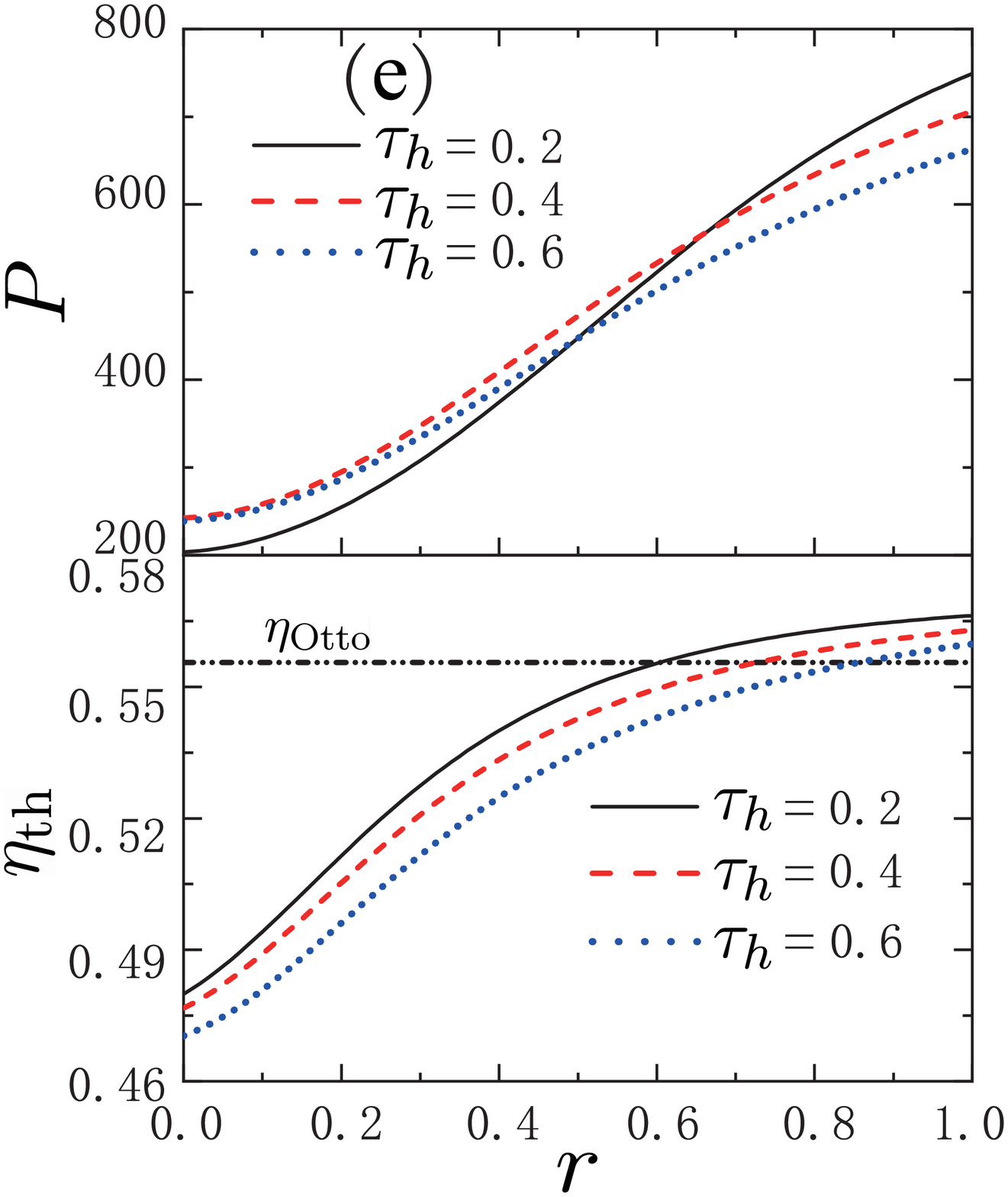}
\caption{{ (a) Transitionless  work, frictional work, and coherent work as a function of  squeezing  parameter $r$ with $\tau_h=0.2, \tau_\mathrm{dri}=5\times10^{-4}$;  (b) Transitionless  work, frictional work, and coherent work as a function of  thermal-contact time with $\tau_\mathrm{dri}=5\times10^{-4}, r=0$}; (c) Power  (up) and  thermodynamic efficiency
 (bottom) as a function of    driving time with $\tau_h=0.5$;   (d) Power (up) and thermodynamic
efficiency (bottom) as a function of  thermal-contact
time with $\tau_\mathrm{dri}=0.0005$. (e) Power (up) and   efficiency (bottom) as a function of squeezing parameter with given $\tau_h$ under $\tau_\mathrm{dri}=0.0005$. In (d), the
dephased engine cycle (labelled for``deph'') are indicated by red
dotted line, blue dashed line, and black  dot-dashed line,
respectively.  The other parameters  are same as Fig. \ref{model}. }\label{p}
\end{figure}

The average work $-\langle \mathrm{w}_{\mathrm{tot}}\rangle$ (\ref{wtot})
 can be split up into the two terms:
$
-\langle
\mathrm{w}_{\mathrm{tot}}\rangle=\langle \mathrm{w}_{\mathrm{deph}}\rangle+\langle \mathrm{w}_{\mathrm{coh}}\rangle $
where $
    \langle \mathrm{w}_{\mathrm{deph}}\rangle
    =\langle
\mathrm{w}_{\mathrm{trls}}\rangle+
    \langle \mathrm{w}_{\mathrm{fri}}\rangle
$, with $\langle
\mathrm{w}_{\mathrm{trls}}\rangle=:\hbar(\omega_{h}-\omega_{c})(\langle n_{t_2}\rangle-\langle n_{t_0}\rangle)$ and $\langle
\mathrm{w}_{\mathrm{fri}}\rangle=:2\hbar\xi(\omega_{c}\langle
n_{t_2}\rangle+\omega_{h}\langle n_{t_0}\rangle)$, is the average
work in the dephasing  case, and $
\langle{\mathrm{w}}_{\mathrm{coh}}\rangle=-2\hbar \omega_{h}\zeta_{ch}-2\hbar \omega_{c}\zeta_{hc}  $
  is the average work associated with quantum coherence.  Here $\langle \mathrm{w}_{\mathrm{trls}}\rangle $ indicates the work in the transitionless case  where the transitions in the instantaneous energy eigenstates are removed,  and $\langle \mathrm{w}_{\mathrm{fri}}\rangle $ represents the additional work that overcomes the inner friction causing unwanted diabatic transitions in instantaneous  energy eigenstates.  { The squeezing results in an increase in the  transitionless work and the coherent work, but a decrease in the amount of frictional work which is always negative, as  shown in Fig. \ref{p}(a).
 The  transitionless work $\langle\mathrm{w_{trls}}\rangle$ increases with increasing thermal-contact time $\tau_h$ to reach the maximum value at which the system approaches to the thermal state, but the effects of $\tau_h$ on both frictional work $\langle\mathrm{w_{fri}}\rangle$ and coherent work $\langle\mathrm{w_{coh}}\rangle$ are particularly small, as shown in  Fig.  \ref{p}(b). The coherent work  displays the oscillations in quick isochoric stroke, since the coherence $C(\rho_{t_2})$,  which interferes  with the coherence generated during the unitary expansion,  is only partially erased.}

{ With consideration of Eqs. (\ref{wtot}) and (\ref{qh}), the
thermodynamic efficiency, $\eta_{\mathrm{th}}=-\langle
\mathrm{w_{tot}}\rangle/\langle{q_h}\rangle$, is then obtained as 
 \begin{equation}
\eta_\mathrm{th}=\eta_\mathrm{Otto}+2(\omega_c/\langle q_h\rangle)[\xi(\langle n_{t_0}\rangle+\langle n_{t_2}\rangle)-\zeta_{hc}-\zeta_{ch}],  \label{etach}
 \end{equation}
 where  $\eta_\mathrm{Otto}=1-\omega_c/\omega_h$ is the so-called Otto efficiency.   Because the times taken for  two isochoric and two unitary strokes are finite, the quantum coherence and  inner friction are created, resulting in that the efficiency depends on  both these two kinds of quantum effects.   Quite interestingly,  the efficiency $\eta_\mathrm{th}$ for the heat engine  ($\langle q_h\rangle>0$)   may surpass  the Otto efficiency $\eta_{\mathrm{Otto}}$ \cite{RJ19}, if  $\xi(\langle n_{t_0}\rangle+\langle n_{t_2}\rangle)>\zeta_{hc}+\zeta_{ch}$. We prove in  SM  Sec. III \cite{sup} that  
 the thermodynamic efficiency,  irrelevant to the Carnot bound \cite{Nie18}, must be bounded by the
generalized Carnot value $\eta_C^{\mathrm{gen}}=1-\beta_h^\mathrm{eff}/\beta_c\!$ with $\beta^\mathrm{eff}_h=\mathrm{ln}\{[2\cosh{2r}+(e^{\beta_h\hbar\omega_h}-1)(\cosh{2r}+1)]/[2\cosh{2r}+(e^{\beta_h\hbar\omega_h}-1)(\cosh{2r}-1)]\}/(\hbar\omega_h)\!$}.

 In contrast to the average work, the average
efficiency of the quantum Otto engine may be ill-defined due to the
possible divergence of the stochastic efficiency \cite{Lut20,Xiao21}. Hence, we resort to large deviation
theory  associated with the exponential decay of probabilities of
large fluctuations, assuming that the quantum engine proceeds in the
long-time limit. The  large deviation function of quantum
efficiency can be given by  \cite{notea}
\begin{equation}
  J(\eta)=-\underset{\varphi_{2}}{\mathrm{min}} \phi(\varphi_{2}\eta,\varphi_{2}). \label{jeta}
\end{equation}
where $\phi(\varphi_{1},\varphi_{2})=\ln\langle e^{\varphi_{1}q_{h}+\varphi_{2}\mathrm{w}_{\mathrm{tot}}}\rangle$.

In Fig. \ref{p}(c), the power oscillates as a
function of the driving  time  $\tau_\mathrm{dri}$, and very quick driving speed results in poor power output.  In our model,  where the driving time $\tau_{\mathrm{dri}}$ is much smaller than the thermal-contact time $\tau_h$ ($\tau_c$) and thus   the total cycle period $\tau_\mathrm{cyc}$ is dominated by $\tau_{h,c}$,
 the contribution of  the driving time to the power mainly comes from  quantum inner friction which  is responsible for irreversible work in each cycle.  The efficiency increases with increasing  driving time, although not monotonically.

 We observe from Fig. \ref{p}(d) that the power first increases in small $\tau_h$ and then decreases with further increase in  $\tau_h$. During
the fast hot isochoric
stroke  the decoherence of the
system is suppressed, yielding the additional,  coherent work $\langle
\mathrm{w}_{\mathrm{coh}}\rangle$ which is responsible for the oscillation. Because the transitionless work $\langle\mathrm{w_{trls}}\rangle$ dominating the total work increases faster
than linearly with increasing $\tau_h$,  the power
increases with increasing $\tau_h$ to a certain maximum value and then
decreases gradually.  The shapes of the efficiency and power curves
are similar, except that $\eta_\mathrm{th}$ increases with $\tau_h$ to reach its
maximum value consistent with $\eta_C^{\mathrm{gen}}$. The oscillations of both the power and the efficiency in Fig. \ref{p}(d) come from the
effect of the dynamical interference between the residual coherence
after the second stroke and the coherence generated in the third
stroke. Interestingly, in the large squeezing case ($r=1$)  leads to  large coherence  [see Fig. \ref{model} (c)] and thus large interference effect, which accounts for large oscillations of these two performance measures, but these two measures   become equivalent to  their respective dephased counterparts  in the long time $\tau_{h}$ where coherence is full erased as they should {[see Fig. \ref{p}(b)]}.
By suitably controlling over the driving and
thermalization times,  the quantum
engine may run in a favorable  regime where both efficiency and power can be enhanced, as shown in  Fig. \ref{p}(d).

The coherent work $\langle \mathrm{w_{coh}}\rangle $ {in Fig. \ref{p}(b)}  may contribute to the
increase of the extracted net work.  If the machine parameters are properly adjusted, the faster
the unitary and thermal-contact processes are performed,
the greater the contribution of the coherence to the
total work extracted, since  coherent work increases with speeding up these processes. The
increase of the extracted work with   shortening of time may lead  to increase power [see  Fig. \ref{p}(e) (up)], and, surprisingly,  
causes efficiency to surpass the Otto limit that is reached when coherence is fully erased or it is not generated along the  unitary stroke. This, the main  message from Fig. \ref{p}(e) (bottom),    confirms  our theoretical prediction (\ref{etach}) that quantum coherence, of purely quantum origin, can lead  to 
a marked difference in machine performance.

{It is of interest to note  the following cases  (see SM Sec. IV \cite{sup})}: (i)  while  the efficiency  is  independent of squeezing in the case of large difference between two reservoir temperatures,   it is sensitively  dependent on the squeezing parameter $r$  in the linear response regime where the difference between two reservoir temperatures is small; (ii) the efficiency depends on the degree of squeezing  in the low-temperature and high-temperature
limits; (iii) In the latter case, by using endoreverisble condition \cite{Cur75} we can reproduce the expression for the efficiency at maximum power \cite{Aba14, Kla17}:  $\eta_{mp}=1-\sqrt{\mathrm{sech}(2r)\beta_h /\beta_c}$.

\begin{figure}[tb]
{\includegraphics[width=8.4cm]{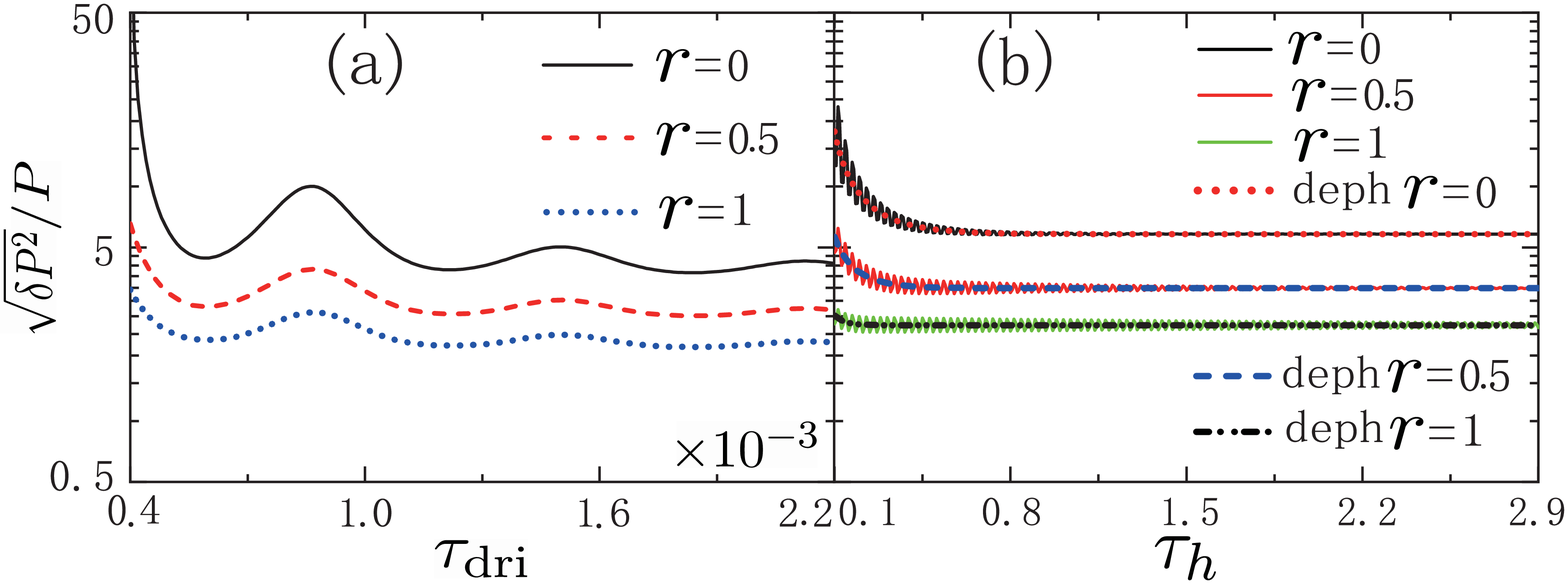}}
\includegraphics[width=8.4cm]{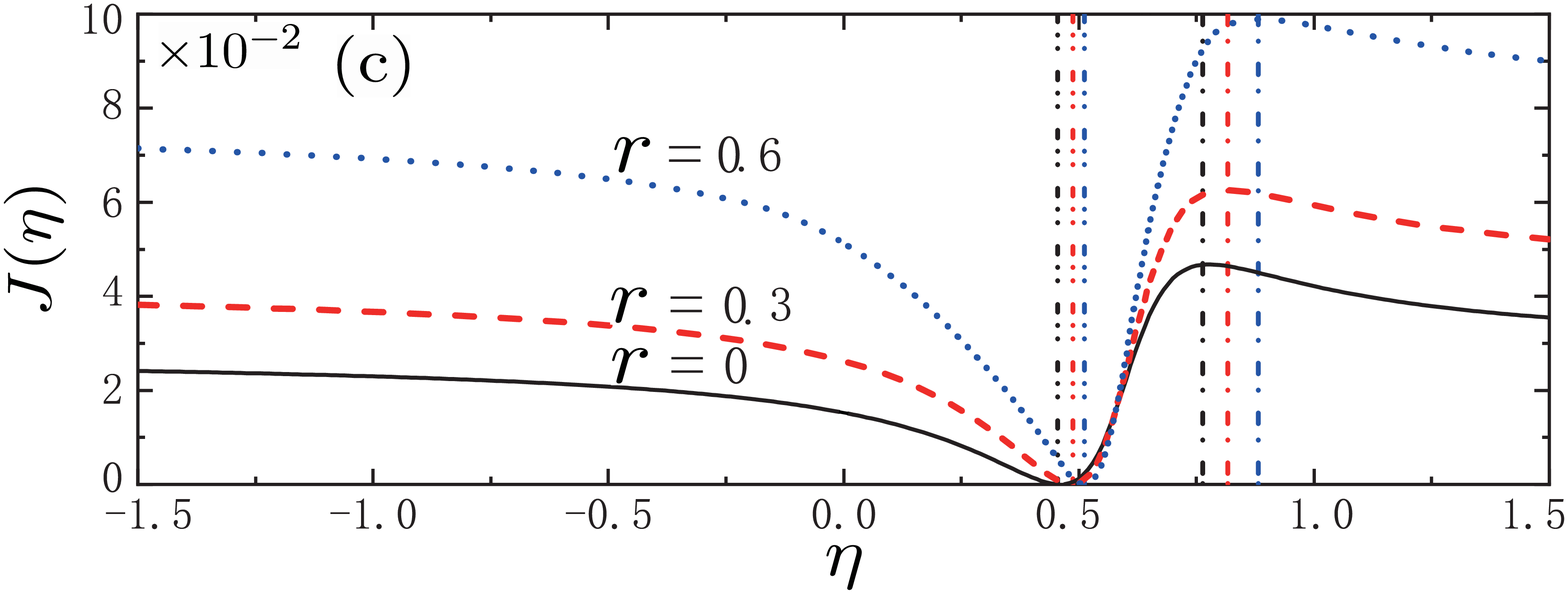}
\caption{
Root-mean-square relative fluctuation of   power, $\sqrt{\delta
P^2}/P$, as functions of (a) driving time $\tau_{\mathrm{dri}}$   and  (b) thermalization time $\tau_h$.  A logarithmic scale is used in the
root-mean-square relative fluctuation (ordinate axis) in (a) and (b).  
(c) Large deviation
function of  efficiency, $J(\eta)$, as a function of stochastic
efficiency $\eta$ \cite{note6}.    In (a)
$\tau_h=0.5$, in (b) $\tau_{\mathrm{dri}}=0.0005$, and in (c) $\tau_\mathrm{dri}=0.001$ and $\tau_h=5$. The other parameters  are same as Fig. \ref{model}.} 
\label{flup}
\end{figure}

The root-mean-square relative fluctuation of   power, $\sqrt{\delta P^2}/P$, which is equivalent to the coefficient of variation of the work, $\sqrt{\delta \mathrm{w^2_{tot}}}/\langle \mathrm{w_{tot}}\rangle$. It  measures the dispersion of the probability distribution and thus describe the machine stability.      The relative power fluctuation  decreases quickly as squeezing paramter $r$ increase, as shown in  Figs. \ref{flup} (a) and  \ref{flup} (b), showing that
reservoir squeezing leads to an increase both  average
work and in work fluctuation, but the increase in the
fluctuation is much less than in the average value. 
 Fig. \ref{flup}(b) shows that the oscillation timescale of power fluctuation with respect to the thermal-contact time $\tau_h$ agrees with the corresponding power and efficiency Fig. \ref{p}(d).  
The relative power fluctuation $\sqrt{\delta P^2}/P$ decreases  while thermal-contact time  $\tau_h$ or driving time $\tau_\mathrm{dri}$ increases.  In physical terms, the larger the  thermal-contact time or driving time (quick driving accounting for quantum coherence) is, the closer the system to the stationary state, so the non-equilibrium thermal fluctuation of the power  is expected to decrease.

  We plot the large deviation function of stochastic efficiency for  the quantum Otto engine in Fig. \ref{flup}(c), where the curve has a maximum when  the stochastic efficiency $\eta=\eta_C^{\mathrm{gen}}$ ($\eta_C^{\mathrm{gen}}$ reduces to $\eta_C$ if $r=0$) and a minimum at $\eta=\eta_\mathrm{th}$.  The function $J(\eta)$  is situated between a maximum at the generalized Carnot
efficiency $\eta_C^{\mathrm{gen}}$ and a minimum at the
thermodynamic efficiency $\eta_{\mathrm{th}}$, which recovers the special
case when squeezing was absent \cite{Ver14, Lut21}. We find that the standard thermodynamic
efficiency is the most likely value, and the generalized Carnot efficiency is the least likely. Furthermore, the
 rate function $J(\eta$) is  strictly larger in presence of squeezing  than the case without squeezing, with the exception of the point $\eta=\eta_\mathrm{th}$.   Figure \ref{flup}(c) shows that the convergence of the heat engine towards  the thermodynamic efficiency is improved by including the reservoir squeezing. This may be understood by noting that quantum efficiency fluctuations, which can be related to machine stability, 
are suppressed under reservoir squeezing.

Experimentally, a quantum Otto engine alternatively driven by a thermal and a squeezed bath  can be implemented by  employing the spin
of the valence electron pertaining  to a single trapped $^{40}$Ca$^+$ ion \cite{Grb19,Yan22} confined in a Paul trap. As the   magnetic
field along $z$ direction yields a Zeeman splitting,
the
Hamiltonian of the  spin  system can be given by
$H=\hbar\omega_z{\sigma_z}/2$. The coupling between  the spin and harmonic motion is mediated via  an optical standing wave   by  a spin-dependent optical dipole force along the oscillation ($x$) direction, which reads  ${\hbar\Delta_{\mathrm{sw}}}\sin(k_{\mathrm{sw} }\hat{x}){\sigma_z}/2$,  with the amplitude of the standing wave $\Delta_{\mathrm{sw}}$ and effective wave number $k_{\mathrm{sw}}$.      The unharmonic term, $\hbar\Delta_{\mathrm{sw}} \sin(k_{\mathrm{sw}}\hat{x})\sigma_z/2$, will be responsible for realization of the squeezed state of the motion \cite{Wu97,Grb19}.  For example, the frequencies of the spin system, determined by magnetic field  which may be along $x$ or $z$ direction,  varies from $2\pi\times 8 $MHz to  $2\pi\times 14.4 $MHz in each  cycle.  The experimental parameters available allow to observe the
machine performance, which confirms our theoretical prediction based on  the choice  of values for
$\omega_c$ and $\omega_h$ falling into a relatively large range (see SM Sec. V  \cite{sup}).

In summary, we have presented a unified thermodynamic theory for a squeezed-bath-driven quantum Otto engine whereby all the variables are periodic  and the efficiency is irrelevant to the Carnot value.  We have shown that the   engine  under squeezing can outperform its non-squeezing counterpart  by
dramatically enhancing efficiency and power output, and even that the efficiency  at positive power may beat the quantum Otto limit.
We have  demonstrated that reservoir squeezing significantly decreases relative power fluctuations and leads to faster convergence of the machine efficiency to its most probable value.
Our findings demonstrate  the potential of quantum engines fueled by nonthermal reservoirs \cite{note2} to realize  ideal nano-scale engines with more efficient, larger power, and higher stability. 

\begin{acknowledgements}  This work is supported by the National Natural Science
Foundation of China (NSFC) (Grants No. 11875034 and No. 61835013),   the National Key R\!$\!\And\!$\!D Program of China under Grants No. 2021YFA1400900, No.2021YFA0718300, and NO. 2021YFA1400243, and  the Major Program of Jiangxi Provincial Natural Science Foundation (Grant No. 20224ACB201007). 
\end{acknowledgements}



\begin{thebibliography}{99}
\bibitem{notea} We  use large deviation theory  to analyze the efficiency statistics. We recall that the large deviation functions of the joint distribution $p(q_h, \mathrm{w_{tot}})$ and the efficiency
distribution $p_{K}(\eta)$ for a large number of cycles ($K\gg1$) are governed by the respective asymptotic forms of $
  p_{K}(q_{h},\mathrm{w}_\mathrm{tot})\asymp e^{-KI(q_{h},\mathrm{w}_\mathrm{tot})},
$
 and 
  $p_{K}(\eta)\asymp e^{-KJ(\eta)}.
$ 
The large deviation functions,  $I(q_{h},\mathrm{w}_\mathrm{tot})$ and $J(\eta)$, describe the exponentially unlikely deviations of  $q_{h}$, $\mathrm{w}_\mathrm{tot}$ and $\eta$ from their most probable values.
The rate function $J(\eta)$ can be obtained from $I(q_{h},\mathrm{w}_\mathrm{tot})$ by the contraction: 
$
  J(\eta)=\underset{q_{h}}{\mathrm{min}}I(q_{h},-\eta q_{h}).
$
Let us define $q_{h}^{(K)}:=\sum^{K}_{j=1}q^{(j)}_{h}/K$ , $\mathrm{w}^{(K)}:=\sum^{K}_{j=1}\mathrm{w}_\mathrm{tot}^{(j)}/K$
$
\phi(\varphi_{1},\varphi_{2})=\underset{K\rightarrow\infty}{\mathrm{lim}}\frac{1}{K}\ln\langle e^{K(\varphi_{1}q_{h}^{(K)}+\varphi_{2}\mathrm{w}^{(K)})}\rangle
 =\ln\langle e^{\varphi_{1}q_{h}+\varphi_{2}\mathrm{w}_\mathrm{tot}}\rangle, $
where we have used $\langle e^{\varphi_{1}q_{h}+\varphi_{2}\mathrm{w}_\mathrm{tot}}\rangle=\int\int dq_{h}d\mathrm{w}_\mathrm{tot}e^{-\varphi_{1}q_{h}-\varphi_{2}\mathrm{w}_\mathrm{tot}}p(q_{h},\mathrm{w}_\mathrm{tot}).
$  Using  the Legendre-Fenchel transform, one  then obtains  the large deviation function of quantum efficiency (\ref{jeta}).
\bibitem{note6}  While  the vertical  dash-double-dot  line patterns (black, red and blue) represent  the values of the stochastic efficiency $\eta$   equivalent to the thermodynamic  efficiencies   ($\eta_{\mathrm{th}}\big|_{r=0}=0.461, \eta_{\mathrm{th}}\big|_{r=0.3}=0.485, \eta_{\mathrm{th}}\big|_{r=0.6}=0.512$), 
  the vertical  dash-dotted lines (black, red and blue) indicate the values of the stochastic efficiency $\eta$   corresponding to the generalized Carnot efficiencies   ($\eta_{C}^{\mathrm{gen}}\big|_{r=0}=0.778, \eta_{C}^{\mathrm{gen}}\big|_{r=0.3}=0.817, \eta_{C}^{\mathrm{gen}}\big|_{r=0.6}=0.884$). \textcolor{blue}{Since the work and heat  are fluctuating quantities, the negative values of the stochastic efficiency $\eta$ with $\eta=\mathrm{w_{tot}}/q_h$ occur when the heat $q_h$ is positive with negative work $\mathrm{w_{tot}}$ or vice versa, and the values of the efficiency larger than $1$ happen when the heat $q_c$ is positive}.
\end{thebibliography}
\end{document}